\documentclass[conference]{IEEEtran}
\IEEEoverridecommandlockouts

\usepackage{cite}
\usepackage{amsmath,amssymb,amsfonts}
\usepackage{algorithmic}
\usepackage{graphicx}
\usepackage{textcomp}
\usepackage{xcolor}
\usepackage{threeparttable}
\usepackage{url}
\usepackage{balance}

\def\BibTeX{{\rm B\kern-.05em{\sc i\kern-.025em b}\kern-.08em
    T\kern-.1667em\lower.7ex\hbox{E}\kern-.125emX}}
\begin{document}
\bstctlcite{IEEEexample:BSTcontrol}

\title{On the Challenges of Fuzzing Techniques via Large Language Models
}

\author{
  \IEEEauthorblockN{
    Linghan~Huang\IEEEauthorrefmark{1},
    Peizhou~Zhao\IEEEauthorrefmark{1},
    Lei~Ma\IEEEauthorrefmark{2},
    Huaming~Chen\IEEEauthorrefmark{1}}
  \IEEEauthorblockA{\IEEEauthorrefmark{1}School of Electrical and Computer Engineering, The University of Sydney, Sydney, Australia\\
  \{lhua5130,\,pzha2332\}@uni.sydney.edu.au, huaming.chen@sydney.edu.au}
  \IEEEauthorblockA{\IEEEauthorrefmark{2}Department of Computer Science, The University of Tokyo, Tokyo, Japan\\
  ma.lei@acm.org}
}


\maketitle

\begin{abstract}
In the modern era where software plays a pivotal role, software security and vulnerability analysis are essential for secure software development. Fuzzing test, as an efficient and traditional software testing method, has been widely adopted across various domains. Meanwhile, the rapid development in Large Language Models (LLMs) has facilitated their application in the field of software testing, demonstrating remarkable performance. As existing fuzzing test techniques are not fully automated and software vulnerabilities continue to evolve, there is a growing interest in leveraging large language models to generate fuzzing test. In this paper, we present a systematic overview of the developments that utilize large language models for the fuzzing test. To our best knowledge, this is the first work that covers the intersection of three areas, including LLMs, fuzzing test, and fuzzing test generated based on LLMs. A statistical analysis and discussion of the literature are conducted by summarizing the state-of-the-art methods up to date of the submission. Our work also investigates the potential for widespread deployment and application of fuzzing test techniques generated by LLMs in the future, highlighting their promise for advancing automated software testing practices.
\end{abstract}


\section{Introduction}
The fuzzing test has been widely employed for software development since the 1990s~\cite{8863940}. Its principle involves generating a series of unexpected or random inputs to assess the reliability and security of software systems. With the evolution of the modern software industry over the years, fuzzing test has become pivotal in modern software testing, playing a critical role in identifying vulnerabilities and ensuring reliable and secure software development. Meanwhile, we have witnessed that large language models have transformed various fields with remarkable performance, including software testing. These novel techniques and cutting-edge knowledge open new possibilities. Using large language models (LLMs), fuzzing testing has gained additional capabilities, enabling the generation of more sophisticated and targeted inputs, enhancing its effectiveness in detecting vulnerabilities in complex software systems.

Software testing generated based on large language models shows a significant improvement in terms of efficiency and accuracy compared to traditional software testing systems. By incorporating the learning and inferencing capabilities of LLMs, these approaches facilitate the automation of software testing processes, including the generation of LLM-based fuzzers for fuzzing tests. Thus, it has become an innovative and trending topic. Several novel methods and frameworks have been proposed to improve fuzzing test systems, such as TitanFuzz~\cite{deng2023large2}, FuzzGPT\cite{deng2023large1}, and other fuzzers tailored to specific software types. Similarly, they all combine different large language models with fuzzing test techniques to develop more robust and efficient fuzzing test systems. In the following sections, we will discuss these approaches in detail, highlighting their key differences and unique contributions of each method to the advancement of fuzzing test systems.

\subsubsection{What's covered?} In this paper, our objective is to provide a comprehensive study covering important works in the field of fuzzing test based on large language models. We thoroughly review the methodologies, benchmarks, metrics, and evaluations, organizing these works in a systematic manner to cover all different aspects of the fuzzing methods. The interaction of large language models in the fuzzing test, how prompt engineering helps large language models complete the fuzzing test tasks, and how to optimize the seed mutation process with the support of large language models are covered with in-depth analysis and discussion. We also provide current challenges and future research directions in the field, capturing the vision for automated testing schema with LLMs and calling for community efforts to build a scalable, human-in-the-loop and sustainable future of testing. 

\subsubsection{Research methodology and research questions} We start with the identified core literature related to LLM-based fuzzer through predetermined criteria, manual screening, and snowballing methods. Our literature review is mainly dedicated to explaining and analyzing all the literature on the combined application of fuzzing test and large language models. We consider a work as relevant strictly referring to any following selection criteria.
\begin{itemize}
\item The work proposes or improves a testing tool, framework, and method for fuzzing tests based on LLMs.
\item The work involves testing methods, concepts, and frameworks of fuzzing test in LLMs environment.
\item The work discusses the relationship between the fuzzing test and LLMs, including theoretical research and methodological discussions.
\item The work discusses the advantages, potential limitations, and future analysis of the combined application of LLMs and the fuzzing test.
\end{itemize}

To thoroughly survey the topic, we propose the following three research questions:

\begin{enumerate}
    \item How do large language models interact with fuzzing test?
    \item What are the advantages of LLM-based fuzzer in comparison with traditional fuzzers? 
    \item What are the future potential and research challenges for LLM-based fuzzers?
\end{enumerate}
In summary, this paper thoroughly reviews the integration of large language models into fuzzing tests, examining methodologies, tools, benefits, and future challenges, with the objective of providing a clear roadmap for advancing automated, intelligent software testing frameworks.

\section{Preliminary}
\label{sec:bk}
\subsection{Large Language Models (LLMs)}
The emergence of large language models has provided great help for different complex language tasks, such as translation, summary, information retrieval, dialogue interaction, etc~\cite{naveed2023comprehensive}. The immense power of the large language model stems from the incorporation of the transformers mechanism into the model's framework, which significantly improves its computational capabilities.

According to statistics by Humza Naveed et al. in July 2023 \cite{naveed2023comprehensive}, there are a total of 75 widely used large language models between 2019 and 2023. Among them, they have categorised these large language models into eight different application fields as following:
\begin{enumerate}
    \item \textbf{General Purpose} – Models designed for versatile applications across diverse contexts, such as GPT-4\cite{openai2024gpt4technicalreport} and DeepSeek-R1\cite{deepseekai2025deepseekr1incentivizingreasoningcapability}.
    \item \textbf{Medical} – Specialized models tailored for healthcare tasks, clinical decision support, and biomedical literature analysis, i.e., Med-PaLM\cite{singhal2022largelanguagemodelsencode} and BioGPT\cite{Luo_2022}.

    \item \textbf{Education} – Models used to enhance learning experiences, personalized tutoring, and educational content generation, including models like Khanmigo\cite{article}.

    \item \textbf{Science} – Models that facilitate scientific research, literature summarization, hypothesis generation, and experiment simulation, such as Galactica\cite{taylor2022galacticalargelanguagemodel}.

    \item \textbf{Maths} – Models optimized for solving mathematical problems, theorem proving, and symbolic manipulation, including Minerva\cite{lewkowycz2022solvingquantitativereasoningproblems}.

    \item \textbf{Finance} – Models specialized for financial forecasting, market analysis, and risk assessment, exemplified by BloombergGPT\cite{wu2023bloomberggptlargelanguagemodel}.

    \item \textbf{Robotics} – Integration of language models with robotic systems for instruction-following, task execution, and environmental interaction, such as RT-DETR\cite{zhao2024detrsbeatyolosrealtime}.

    \item \textbf{Coding} – Models optimized for software development tasks, automated code generation, debugging, and technical documentation, notably Codex\cite{chen2021evaluatinglargelanguagemodels} and Code Llama\cite{rozière2024codellamaopenfoundation}.
\end{enumerate}

In addition, the model framework can be divided into three categories: decoder-only language model, encoder-only masked language model and encoder-decoder language model~\cite{fu2023decoderonly}. The decoder-only language model can perform zero-shot program synthesis by generating output in a left-to-right fashion\cite{fu2023decoderonlyencoderdecoderinterpretinglanguage}. For the statistics of this review, we observe that hybrid technology of LLMs and fuzzing test predominantly uses decoder-only language models, such as InCoder~\cite{fried2023incoder}, CodeX~\cite{chen2021evaluating}.

On the other hand, for encoder-decoder model, such as CodeGen~\cite{nijkamp2023codegen}, it is presented by Salesforce Research to handle natural language processing and program generation tasks. The CodeGen model combines an encoder and decoder architecture, which makes it particularly effective at understanding and generating text, including programming languages.

Encoder-decoder models exhibit superior capabilities to handle complex language understanding and generation tasks compared to decoder-only or encoder-only models. For example, CodeGen can not only generates code, but also understand and process more complex programming language structures and logic. It is especially well-suited for tasks that require a deep understanding of code semantics and structure, such as code translation, code summarization, and code generation.

\subsection{Fuzzing Test}
The first fuzzing tool was designed by Miller et al. (1990) to test the reliability of the software and the system~\cite{CHEN2018118}. Ever since its introduction in the early 1990s, fuzzing has become a widely used technique to test software correctness and dependability. Before 2005, this technique was still in the early stage of development, which are mostly black-box fuzzing using random mutation. From 2006 to 2010, some fuzzing systems adopted taint analysis techniques and symbolic execution-based fuzzing was adequately developed. Several fuzzing test systems are published, some of which are even incorporated in the commercial product. Between 2011 and 2015, we witnessed the evolution of the fuzzing test characterized by three main characteristics\cite{10.1145/3512345}. Firstly, coverage-guided fuzzing has become an important topic for both academic and industrial domains. Secondly, different types of scheduling algorithms are used in fuzzing processes. Third, improving the efficiency of fuzzing by integrating various techniques has been a critical aspect of testing. In 2016-2017, many fuzzing systems are improved based on AFL (American fuzzy loop)~\cite{coredumpAmericanFuzzy}, such as AFLFast~\cite{bohme2016coverage} and AFLGo~\cite{bohme2017directed}. Since then, there has been a trend of combining multiple techniques for better fuzzing test techniques, during which the concept of machine learning is introduced to fuzzing. Generally, it can be categorized into two main types, mutation-based fuzzing test and generation-based fuzzing test\cite{li2018fuzzing}. These two types of fuzzing test can be used for different types of software. 

According to research on the art, science and engineering of fuzzing\cite{li2018fuzzing}, current fuzzer, a program that performs fuzzing test on a program under test (PUT), can be separated into three classes, \textbf{black-box fuzzer}, \textbf{white-box fuzzer} and \textbf{grey-box fuzzer} \cite{khan2012comparative}. The traditional black-box fuzzing test technique does not require prior information about the target PUT. It employs pre-defined mutation rules to mutate seed files and generate new erroneous inputs. In contrast, a white-box fuzzing test necessitates the use of PUT information to guide the generation of test cases. Theoretically, the white-box fuzzing test has the potential to cover every path of the program. Grey-box fuzzing test falls between black-box and white-box fuzzing test. It generates effective test cases with partial information of the PUT. A typical approach involves obtaining the code coverage of the PUT during runtime and leveraging this information to adjust the mutation strategies~\cite{8371326}.

\begin{figure*}[h]
    \centering
    \includegraphics[width=\textwidth]{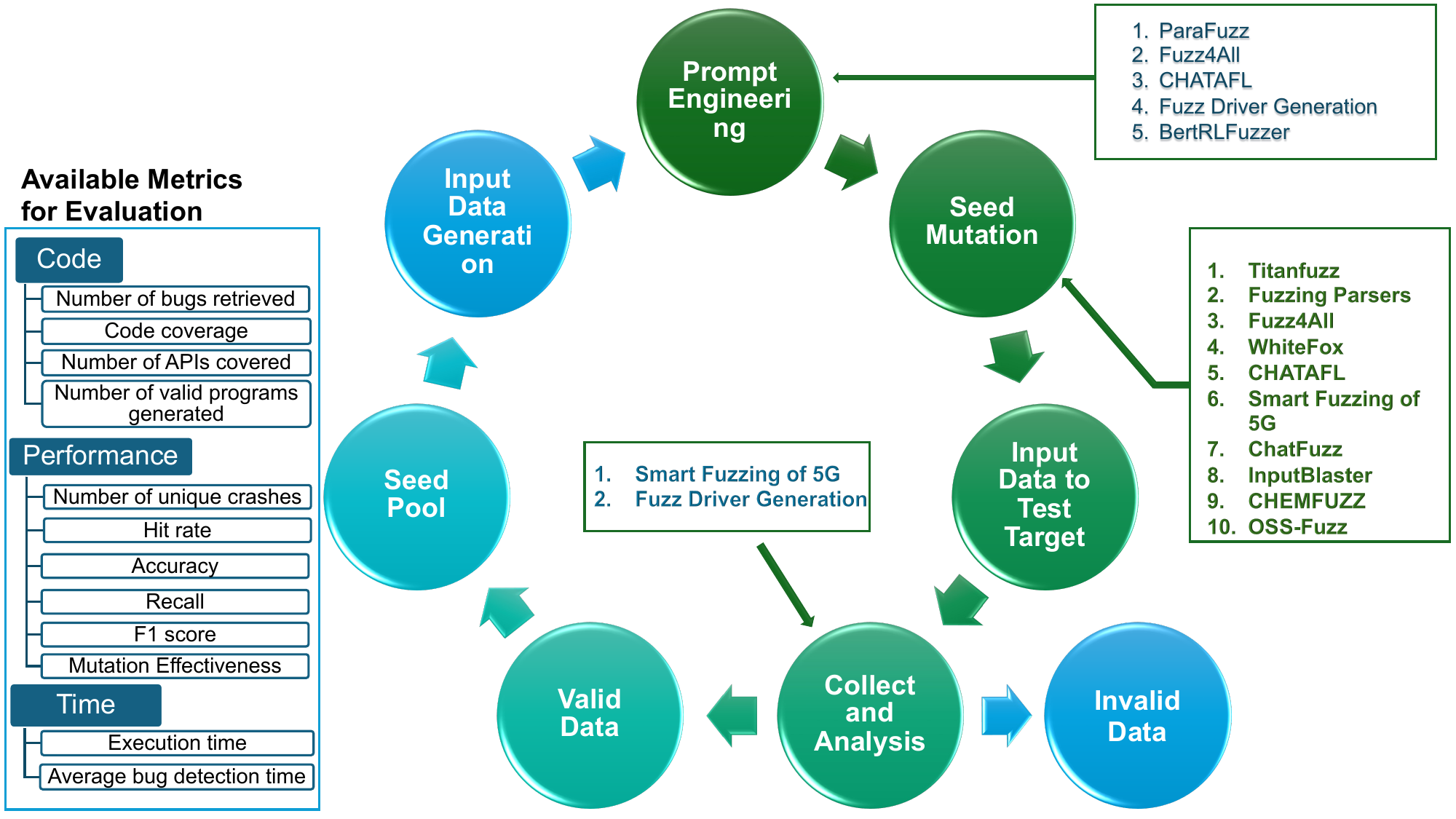}
    \caption{A General overview of LLM-based Fuzzer}
    \label{fig:llm}
\end{figure*}

\section{LLM-Based Fuzzing Test: Where We are?}
The scope of the application of the fuzzing test is wide, and it has been implemented in various domains of the software industry to assess the reliability and security of the software system. However, the traditional fuzzing test has presented certain limitations, encouraging researchers to explore the integration of emerging large language models with the fuzzing test to improve its efficiency and accuracy.

In this paper, we summarize two significant research topics which have been significantly investigated recently in the literature, which are described as follows:
\begin{itemize}
    \item \textbf{Fuzzer by LLM:} This research topic explores the possibility and methods of introducing LLMs into the traditional fuzzing test process. By combining LLM with the fuzzing test, this method aims to enhance the functionality and performance of the traditional fuzzing test, thus improving test coverage and defect discovery efficiency and ultimately achieving better testing results. Through a thorough analysis of LLM-based fuzzers, as shown in Figure~\ref{fig:llm}, we find that existing LLM-based fuzzers usually introduce LLMs into prompt engineering and seed mutation to enhance the performance of fuzzing test\cite{10.1145/3663529.3663784}.
    \item \textbf{Fine-Tuning Fuzzer:} This research topic is more specialized, focusing on selecting and fine-tuning an open source large language model as a fuzzer that can automatically generate test code snippets in batches. This approach aims to optimize and improve the efficiency and accuracy of the fuzzing test through the automated application of the fine-tuned LLM model in the test seed generation process.
\end{itemize}

In addition, we summarize the most commonly used metrics based on all technologies to evaluate the performance of LLM-based fuzzers. They can be generally categorized into three types, the metrics related to code, performance and time:
\begin{itemize}
    \item \textbf{For the code-related metrics}, the number of vulnerabilities detected is one of the key criteria to evaluate the performance of each LLM fuzzer and usually represents the main contribution of the research. For the evaluation of the code coverage, the metric provided by \underline{ProFuzzBench} \cite{10.1145/3460319.3469077} is highly representative, and CHATAFL \cite{chatafl} uses this metric to detect the code coverage of its fuzzer.
    
    \item \textbf{For the performance-related metrics}, \underline{the hit rate} is a commonly used metric. It refers to the efficiency of inputs generated by the fuzzer to target test objectives. Mutation effectiveness \cite{gopinath2022mutationanalysisansweringfuzzing} is closely related to the mutation score and serves as a standard to measure the quality of seed mutation. It is worth mentioning that some techniques also use \underline{the F1 Score} \cite{sokolova2006beyond} as a metric. It is calculated using precision and recall and ranges from 0 to 1, with a higher value indicating better performance in terms of both precision and recall.
    
    \item \textbf{For the time-related metrics}, the execution time represents the speed at which the fuzzer runs. On the other hand, average detection time is a standard measure used to gauge the fuzzer's ability to find vulnerabilities in the test objectives. 
\end{itemize}
As shown in Table~\ref{tab:plain}, we classify all LLM-based fuzzers based on LLM models, benchmarks, and testing types, and conduct classification discussions and future work discussions on the literature in later sections.

\subsection{Fuzzer by LLM: Prompt engineering and Seed Mutation}
The core of prompt design is to formulate a series of guiding prompts to help LLMs better understand input and output tasks~\cite{yan2023parafuzz}. In LLM fuzzer, by using prompt in the seed generation stage, complex and lengthy inputs can be refined to efficient test information, which can significantly improve the performance and convenience of the fuzzing test. Specifically, we can use the prompt to clearly indicate the specific tasks that the large language model needs to handle. For example, in Fuzz4All \cite{xia2024fuzz4all}, the automatic prompt generation technology (Autoprompting) is used to refine user input and generate more concise and effective information prompts by refining LLM~\cite{zhang2023understanding}.

We constrain the output of the large language model by providing relevant prompts such as \textbf{`task description'}, \textbf{`expected results'}, \textbf{`responsibilities'} and \textbf{`role statement'}, \cite{chen2024unleashingpotentialpromptengineering}, \cite{meng2024large}, \cite{jha2023bertrlfuzzer}, \cite{10.1145/3605157.3605173} to obtain more objective and accurate results. This process can help the fuzzing tester generate higher quality test seeds. For example, Titanfuzz uses prompt engineering to help the large model better understand the data input, and generates a group of test code seeds by setting a series of continuous task prompts.

In actual operation, the application of prompt engineering in LLM fuzzer can include the following aspects:
\begin{itemize}
    \item \textbf{Simplifying lengthy and complex information into refined test inputs.} In the seed generation process, prompts are used to control output quality to ensure the generation of concise and effective test code seeds. \cite{xia2024fuzz4all} explicitly proposes that before the large language model generates output, the user's redundant input is refined through the use of prompts to ensure the accuracy and effectiveness of the generated results.
    
    \item \textbf{Generating test inputs for specified special scenarios.} With appropriate prompts, large language models can be guided to generate seed codes for testing boundary cases or other special scenarios. \cite{qiuchemfuzz} clearly states that prompts can guide large language models to generate outputs that meet the requirements of specific tasks and ensure that the generated content follows the direction and output template set by the prompts.
    
    \item \textbf{Mutating testing seeds.} In fuzzing test, prompts can be used to perform seed mutation operations. Compared with traditional seed mutation methods, modern methods tend to use prompts to guide large language models to perform specified mutation operations, thereby converting original testing seeds into mutated seeds. This method significantly improves the efficiency of seed mutation in fuzzing test. \cite{yang2023whitebox} mentioned that prompts are used through feedback loops, such as \textit{`Please generate different valid [TARGET INPUT] example with [INPUT SPECIFICATION] meets the specified requirements.'}, to regenerate mutated test seeds, thereby optimizing the mutation process.
\end{itemize}

Among traditional mutation fuzzers, seed mutation is the most important step as most model-free seed fuzzers generate a seed which is an input to the PUT. The purpose of modifying the seed is to generate new test cases that are mostly valid but also contain outliers to trigger the crash of PUT.
Traditional seed mutation technologies include Bit-Flipping~\cite{vdalabsWelcome}, Block-Based Mutation~\cite{8811923}, Dictionary-Based Mutation~\cite{githubGitHubGooglehonggfuzz}. However, seed mutation based on large language models has brought significant advantages to the performance of fuzzing test technology. It can understand and modify these data types through context awareness and natural language processing capabilities. It can maintain the validity of the file format while performing more complex mutation operations, such as rewriting code fragments, changing data structures, and even simulating complex user interaction~\cite{googleblogAIPoweredFuzzing}, \cite{wu2023smart}.

\cite{yan2023parafuzz} mentioned that identifying synonyms and intersections of content words and function words is a challenge in the mutation process. Synonyms are needed to retain the meaning of the statement unchanged during the mutation process. Therefore, a Meta prompt method is presented to solve this problem by using ChatGPT to perform mutation operations. The specific operation is that the GPT model performs statement mutation operations based on a set of special prompts. Thus, data can retain their original meaning while being reorganized by other data. Furthermore, Titanfuzz used CodeX and specific prompts to generate large amounts of initial input data. Second, to obtain more diverse input data, they masked certain parts of the initial input data and then used fill-in models to fill in these masked parts. This enables the production of richer and more varied input data sets. Titanfuzz employs four methods to mask the content of the seed and then mutate it through CodeGen, namely parameter mutation operator, prefix/suffix mutation operator, and method mutation operator. Consequently, the program will mark the SPAN symbol at a specific position of the seed to prompt the model to randomly generate new content for effective seed mutation operations.

\subsection{Fine-Tuning Fuzzer}
Another important category of LLM-based fuzzers is to prime LLMs as the core fuzzer for bug triggering. It takes advantage of the automation and learning capabilities through LLMs, which cover various learning processes. At the time of this survey, FuzzGPT is the state-of-the-art technique based on LLM using the fine-tuning method. In detail, it mainly uses advanced generative artificial intelligence models such as CodeX and CodeGen. CodeX, as the core LLM, fully demonstrates the potential of large language models in fuzzing tests, while CodeGen is mainly used to optimise fine-tuning strategies. \textbf{The core of a fine-tuning fuzzer is to guide large language models to generate abnormal inputs through different learning modes and fine-tuning methods, so as to effectively fuzz test deep learning libraries which have improved the state of the art across different domains} \cite{pandey2019comprehensive}. Since the fine-tuning fuzzer involves the learning and fine-tuning process of large language models, the selection and processing of data are crucial \cite{xia2024understandingperformanceestimatingcost}. This technique is based on the key assumption that historical vulnerability triggers may contain valuable information and code snippets for vulnerability discovery. The fine-tuning fuzzer relies on these historical datasets to achieve effective contextual learning or fine-tuning of LLM.

Code snippets of historical vulnerability triggers from DL libraries and databases will be recorded into the data sets. Subsequently, with these datasets, appropriate methods are determined to perform contextual learning or fine-tuning of LLMs. The following sections will explain the details about three different methodologies.

\subsubsection{Few-shot learning} Few-shot learning aims to use previous knowledge and examples to rapidly generalize to new tasks \cite{brown2020language}, \cite{wang2020generalizing}. In FuzzGPT few-shot learning, a certain number of vulnerability-triggering code snippets are provided as examples to the LLM. The format of the examples usually includes the API name, the vulnerability description, and the code snippet that triggers the vulnerability. By learning these examples, the LLM is able to generate output formats that meet the requirements without additional transformation. In addition, the LLM can also generate similar edge case code snippets by observing these code snippets, thereby completing the task without modifying the model parameters. It is worth mentioning that chain-of-thought prompts are an indispensable part of few-shot learning. Such prompts do not require the model to generate output directly, but guide the model to complete the task step by step. For example, the name of the test target API is included in the input example. The model needs to generate a description of possible vulnerabilities in the format of the input example and then generate code snippets that may trigger specific vulnerabilities.

\subsubsection{Zero-shot learning} The main feature of zero-shot learning is to recognize and understand the new concepts by having a description of them~\cite{petroni2019language}, \cite{pmlr-v37-romera-paredes15}. In FuzzGPT, zero-shot learning includes two variants. The first is zero-shot completion, in which only partial code snippet is provided in the LLMs input example. The code snippet is selected and created from the historical vulnerability-triggering code dataset. Before entering into LLM, a natural language prompt, such as `\# The following code reveals a bug in target-api', is first required to be added so that LLMs understands the test target and allows technicians to apply zero-shot learning to any target API. Next, code snippets are selected from the dataset and the suffixes of some code snippets are randomly removed, leaving the prefix. The new code snippet is input to LLMs as an example, and LLMs needs to generate a complete code snippet as output. The second is zero-shot editing. While the sample code snippet is not partially removed, LLMs is prompted to edit the code snippet through natural language comments (such as \textquotedblleft\# Edit the code to use target-api\textquotedblright) and reuse most of the original code snippet in the example. The zero-shot editing method allows the model to fully utilize historical vulnerability-triggering code to generate brand new fuzzing test output.

\subsubsection{Fine-tuning} For this methodology, some potential factors could affect the performance of the model including pre-training condition, fine-tuning condition, fine-tuning data size and method~\cite{zhang2024scalingmeetsllmfinetuning}. In FuzzGPT, contextual learning only uses pre-trained models, while fine-tuning retrains the model by exploiting historical vulnerability trigger datasets and directly modifies model parameters. The training sample format is consistent with the examples in few-shot learning. Starting from the pre-trained LLMs, the model weights are continuously adjusted and updated through each training generation result. The magnitude of the weight adjustment is defined by the loss function of fine tuning. Each DL library has its own fine-tuning model, which uses its own historical vulnerability trigger code snippets for learning, so that it has the ability to identify different types of vulnerabilities. In addition, the input method of the fine-tuning model is the same as that of few-shot learning, creating prompts based on specific target APIs.

\begin{table*}[t]
    \centering
    \renewcommand{\arraystretch}{1.1} 
    \setlength{\tabcolsep}{4pt} 
\resizebox{\textwidth}{!}{%
    \begin{threeparttable}[b]
    \caption{Overview of core techniques in the important literature}
    \label{tab:plain}
    \begin{tabular}{p{3.5cm}p{3cm}p{3.5cm}p{5.5cm}p{1.6cm}}
        \hline
        Methods  & Domain classification & Models &Benchmarking Dataset & Test type \\
        \hline
        TitanFuzz\tnote{1}     & AI software     &  CodeGen   &Pytorch, TensorFlow   &Black-Box\\
        FuzzGPT\tnote{2}      & AI software     & CodeX, CodeGen   &Pytorch, TensorFlow   &Black-Box\\
        ParaFuzz     & AI software     & ChatGPT   &Amazon Reviews, SST-2, IMDB, AGNews   &Black-Box\\
        Fuzzing Parsers     & AI software     & OpenAI's GPT-4   &Tomita grammars, other novel grammars   &Black-Box\\
        BertRLFuzzer & AI software     & BERT, Reinforcement Learning   &victim websites, Other fuzzing test tools' benchmark   &Grey-Box\\
        Fuzzing BusyBox\tnote{3} & AI software & GPT-4 & BusyBox ELFs proprietary firmware dataset & Black-Box\\
        Fuzz Driver Generation\tnote{4}     & Non-AI software     & GPT-3.5, GPT-4   &NAIVE-K, BACTX-K   &Black-Box\\
        Fuzz4All\tnote{5}     & Non-AI software     & GPT-4, StarCoder   &Six input languages and nine systems (SUTs)   &Black-Box\\
        WhiteBox\tnote{6} & Non-AI software & GPT-4, StarCoder & PyTorch Inductor, TensorFlow Lite, TensorFlow-XLA, LLVM & White-Box\\
        OSS-Fuzz\tnote{7}     & Non-AI software     & Google LLM   &tinyxml2, OpenSSL   &Black-Box\\
        CHATAFL\tnote{8}      & Non-AI software     & GPT-3.5 turbo   &PROFUZZBENCH   &Grey-Box\\
        InputBlaster     & Non-AI software     & ChatGPT, UIAutomator   &18 baselines from various aspects   &Black-Box\\
        Smart Fuzzing of 5G  & Non-AI software     & Google Bard, ChatGPT   &OAI5G configuration file   &Black-Box\\
        CHATFUZZ     & Non-AI software     & GPT-3.5 turbo  &Unibench, Fuzzbench, LAVA-M   &Grey-Box\\
        CHEMFUZZ    & Non-AI software     & GPT-3.5, claude-2  &Siesta(Quantum chemistry software)   &Grey-Box\\
        \hline
    \end{tabular}
    \begin{tablenotes}
        \item[1] \url{https://github.com/ise-uiuc/TitanFuzz}
        \item[2] \url{https://github.com/ise-uiuc/FuzzGPT}
        \item[3] \url{https://github.com/asmitaj08/FuzzingBusyBox_LLM}
        \item[4] \url{https://sites.google.com/view/llm4fdg/home}
        \item[5] \url{https://zenodo.org/records/10456883}
        \item[6] \url{https://github.com/ise-uiuc/WhiteFox}
        \item[7] \url{https://github.com/google/oss-fuzz}
        \item[8] \url{https://github.com/ChatAFLndss/ChatAFL}
        
    \end{tablenotes}
    \end{threeparttable}
    }
\end{table*}

\section{LLM-Based vs. Traditional Fuzzers: A Comparative Analysis}
Compared to traditional fuzzers, LLM-based fuzzers have the following advantages:

\subsection{Higher API and Code coverage:}
By comparing TitanFuzz with current state-of-the-art API-level (such as FreeFuzz~\cite{wei2022free} and DeepREL~\cite{deng2022fuzzing}) and model-level (such as LEMON~\cite{wang2020deep} and Muffin~\cite{gu2022muffin}) fuzzers, it was found that TitanFuzz's API coverage in TensorFlow and PyTorch increased by 91.11\% and 24.09\% respectively. As model-level fuzzers, LEMON and Muffin use a small part of the hierarchical API (such as Conv2d), so their coverage is low. TitanFuzz can generate arbitrary code by combining generative (Codex) and infill (InCoder) large language models (LLMs) to achieve optimal API coverage.

CHATAFL, as an LLM-based fuzzer for network protocols, exhibits higher average code coverage compared to other traditional fuzz testers of the same kind without LLMs, such as AFLNET~\cite{pham2020aflnet} and NSFuzz~\cite{qin2023nsfuzz}. Compared to AFLNET, CHATAFL achieves an average of 5.8\% more branch coverage. Unlike NSFuzz, the coverage increases to 6.7\%. This indicates that CHATAFL has a wider detection range and a higher chance of discovering unknown bugs.
 
\subsection{Higher computational efficiency:}
In terms of overall code, TitanFuzz achieved 20.98\% and 39.97\% code coverage on PyTorch and TensorFlow respectively, significantly exceeding DeepREL~\cite{deng2022fuzzingdeeplearninglibrariesautomated} and Muffin. Compared with DeepREL, TitanFuzz's code coverage in PyTorch and TensorFlow increased by 50.84\% and 30.38\%, respectively. Although TitanFuzz has a higher time cost, using only the seed generation function and testing against the APIs covered by DeepREL is significantly better than DeepREL and takes less time, which shows the advantage of directly using LLMs to generate high-quality seeds.
 
\subsection{Capability to detect more complex errors:}
Traditional fuzzers typically generate test cases randomly based on prescribed rules or methods. Such methods lack a deep understanding of the code structure, logic, and context. Therefore, they may not be effective in exploring complex programming patterns or identifying advanced vulnerabilities. Additionally, traditional fuzzers do not explicitly leverage historical data or program patterns. In contrast, LLMs can learn from extensive historical code and bugs, thereby mining latent bug patterns for novel vulnerability discovery. In FuzzGPT, a total of 76 bugs were detected, 61 of which were confirmed, including 49 confirmed as previously unknown bugs (6 of which have been fixed). In particular, FuzzGPT detected 11 new high-priority bugs or security vulnerabilities, highlighting that LLM-based fuzzers can find deeper programming vulnerabilities.

For example, when testing the same target using CHATAFL, NSFUZZ\cite{10.1145/3580598}, and AFLNET, with the same number of runs and time, CHATAFL discovered 9 new vulnerabilities. In contrast, NSFUZZ only found 4 vulnerabilities, while AFLNET found three. Fuzz4All demonstrates descent performance, being able to discover 76 bugs in widely used systems such as GCC, Clang, OpenJDK, and 47 of these bugs have been confirmed as previously unknown vulnerabilities.

\subsection{Increased automation:}
Traditional fuzzing test requires a significant amount of time, effort, and manual labor. Seed generation is an indispensable part of the fuzzing test, and creating diverse and effective inputs requires a lot of manual work and expertise. Additionally, using existing seeds to generate new input through mutation is a time consuming process. Automating the fuzzing test with LLMs can effectively address these issues. InputBlaster ~\cite{liu2023testing} leverages the automation capabilities of LLMs to generate high-quality seeds based on input prompts. After each test, it mutates the seeds according to different prompts, enabling the generation of new seeds for further testing. Implementing an automated fuzzing test can save considerable costs and is expected to be a prominent trend in the future.

\section{Challenges and Future Directions}
As mentioned above, LLM-based fuzzers can be classified into two types. The first one is to find the historical data set of the object being tested, classify the historical error code fragments and vulnerabilities and provide them to the model for learning, and finally train a specialized fuzzing test model to test the object.  

Another section focuses on introducing LLMs into specific steps in the traditional fuzzing process to improve the fuzzer performance. For example, TitanFuzz mentions using CodeX and CodeGen to change the seed mutation process in traditional fuzzing test to improve code coverage and test success rate. On the other hand, FuzzGPT learns from a historical dataset that contains bugs and vulnerabilities for the deep learning system. However, the performance comparison between \textbf{FuzzGPT(Fine-Tuning Fuzzer)} and \textbf{Titanfuzz(Fuzzer by LLM)} shows that FuzzGPT is better in terms of code coverage and efficiency. 

To define a good fuzzer, a set of criteria is discussed in~\cite{mallissery2023demystify}, including: 
\begin{enumerate}
    \item Able to detect all vulnerabilities of the test targets
    \item Performing in-depth analysis of multiple targets when detecting code-level vulnerabilities from the interaction between multiple targets
    \item A fuzzer should identify different types of bugs
\end{enumerate}
Therefore, we anticipate that, in the future development of the LLM-based fuzzer, a more promising approach would involve allowing the model to learn from historical data and evolve into a professional fuzzer. LLMs can better understand the intricacies of software code and bugs, thus aligning with the comprehensive criteria, rather than altering the operation process of the traditional fuzzer. 

\subsection{LLMs hallucinations}
The application of Large Language Models in fuzz testing has significant potential but faces substantial challenges due to hallucinations, a phenomenon where LLMs generate plausible, yet incorrect, outputs without adequate contextual information. Hallucinations in fuzzing drivers manifest in various forms, such as type mismatches, improper initialization of function arguments, and the usage of non-existent functions or identifiers, contributing to high false-positive rates and decreased test accuracy \cite{zhang2025llmhallucinationspracticalcode}. A recent evaluation in the OSS-Fuzz project highlighted that only approximately 40\% of the GPT-4 synthesized drivers were error-free, the remainder showing hallucination-induced inaccuracies \cite{jiang2024fuzzingmeetsllmschallenges}. Moreover, due to constraints in training data, LLMs are not able to handle unseen APIs accurately, resulting in incorrect invocation sequences and limited testing coverage \cite{10.1109/TSE.2024.3368208}. Addressing hallucinations is therefore crucial for improving the effectiveness of LLM-assisted fuzzing.

\subsection{Agents}
LLM Agents\cite{Wang_2024} \cite{jin2024llmsllmbasedagentssoftware} are comprehensive intelligent agents that integrate multi-model collaboration to automate specific tasks. To improve the efficiency and accuracy of automated testing, LLM agents generate higher quality and more efficient test seeds through multi-model collaboration, and dynamically adjust inputs during the test process to more effectively trigger potential vulnerabilities. In LLM agents, prompt engineering plays a key role, helping LLM agents automatically generate input data or adjust adaptive input strategies based on context. Specifically, in fuzzing test applications, LLM agents can identify specific patterns or contextual contexts of code to generate more targeted inputs to efficiently trigger potential errors or vulnerabilities. In addition, through repeated iterations and reinforcement learning\cite{SHAKYA2023120495}, LLM agents can gradually improve the coverage and accuracy of fuzzing test while continuously optimizing the quality of test cases. In the future, LLM agents may further combine reinforcement learning algorithms to achieve more advanced adaptive optimization of test case generation and testing processes, and ultimately move towards a fully automated and efficient fuzzing test process.

\subsection{Pre-training Data}
In the context mentioned above, training specialized models for the fuzzing test represents a prospective research direction. However, challenges about the sufficiency and quality of pre-training data are notable.~\cite{lin2020identifying} indicate that biases in datasets can arise, potentially reflecting societal inequalities or the prejudiced perspectives of data scientists. This suggests that the quality of pre-training datasets used for models can be skewed.

~\cite{kaddour2023challenges} also point out that the repetitiveness and scale of pre-training datasets can impact the LLMs performance, such as the presence of semantically near-duplicate content which could increase the model's dependency on the data. LLM-based fuzzers may encounter these issues. When the volume of data is inadequate or of poor quality, LLMs may not achieve exceptional performance in specific domains. Whether LLM-based fuzzers can surpass traditional fuzzers in terms of performance remains to be observed. Addressing these challenges will be essential for the future success of LLM-based fuzzers.

\subsection{Computational Efficiency}
LLMs are highly complex models that require substantial computing resources to generate code. This complex calculation process requires a large amount of time and might be highly expensive in economic terms~\cite{liu2024optimizingllmqueriesrelational}.  For instance, TitanFuzz not only generates code but also generates high-quality, meaningful program snippets, which may involve multiple iterations and adjustments to ensure that the generated program meets specific quality standards. Additionally, to generate unique and diverse programs, additional checks and filters are required to prevent duplication of existing code snippets. Furthermore, in the context of deep learning libraries, API calls may involve complex data structures and algorithms. Generating code that effectively calls these APIs and triggers potential errors may require more complex logic and extended processing time. On the other hand, \cite{li_2023} noted that as the demand for LLM data continues to increase, the computing cost, time, and hardware requirements are growing exponentially and may eventually peak by 2026. The computational cost is estimated to be 20\% of US GDP. This trend will bring greater challenges to the development of LLMs, especially when computing resources are limited.

\subsection{Benchmark for LLM-Based Fuzzing Test}
The criteria for evaluation are mainly customized for different testing objectives. The ultimate measure of a fuzzer is the number of distinct vulnerabilities identified. In addition, it is common to assess fuzzer performance based on code coverage~\cite{klees2018evaluating}. Many techniques measure code coverage, and some rely solely on code coverage as the evaluation criterion, such as FairFuzz~\cite{9000039}. 

Fuzz4All mentioned that the code coverage measurement was used to evaluate the performance of the fuzzer. To maintain consistency, the line coverage for each evaluation target was reported. However, there is no inherent basis to directly associate maximizing code coverage with vulnerability identification. In Kless's fuzzing test evaluating research, a statistical analysis of 32 fuzzy testing papers is conducted to examine the experimental methods. They emphasize an urgent need for well-designed and thoroughly evaluated benchmark suites for fuzzing tests.

Many traditional fuzzing test evaluation criteria have been used to evaluate LLM-based fuzzer. However, a proper evaluation for the specific fuzzing techniques is still hard to achieve. \cite{metzman2021fuzzbench} Consequentely, there is a need for a universal and tailored evaluation framework specifically designed for LLM-based fuzzer. The new evaluation framework should extend the evaluations of different LLMs used by fuzzers. It should compare the efficiency of LLM-based fuzzers with traditional fuzzers under consistent parameters, environments, and test subjects. Thus, it could be universally applicable to the majority LLM-based fuzzers. It will be a crucial standard for measuring the performance of such technologies in the future.

\subsection{Full Automation Fuzzing Test}
LLM-based fuzzers significantly reduce manual labor compared to traditional fuzzers. ChatFuzz~\cite{hu2023augmentinggreyboxfuzzinggenerative} leverages prompt engineering from LLMs for seed selection and mutation. Fuzz4All updates the LLM prompts after each iteration to avoid generating the same test inputs and to generate higher quality inputs. 

Similar work in traditional fuzzing tests requires the intervention of a developer. In Google's work, it is mentioned that manually copying results after testing a project with the traditional fuzzer OSS-Fuzz consumes a considerable amount of time, which is notably improved by integrating LLM with OSS-Fuzz. These techniques utilize the characteristics of large language models to some extent, making the developer's workload lighter. Currently, LLM-based fuzzers cannot achieve fully automated fuzzing test generation, but this is undoubtedly the direction of future development.

\subsection{Potential direction on hardware testing}
According to our investigation and research, current LLM fuzzing test research in the hardware field is still in its early stages. Exploring the application potential of fuzzing test methods based on large language models (LLM) in hardware testing has gradually become a focus of research attention~\cite{saravanan2024emergencehardwarefuzzingcritical}. The research results show that the LLM fuzzing test can be used as an innovative auxiliary tool to support the testing of hardware communication transmission and component interaction. Although there are still some technical challenges in the application, through reasonable model optimization and integration with the hardware test environment, the LLM fuzzing test shows certain application potential in the hardware design and verification process. For example, the LLM fuzzing test can generate test inputs with different formats and contents to simulate communication data packets and detect whether they will cause abnormal conditions in the protocol stack. In the early stage of hardware design, the LLM fuzzing test generates diverse test inputs in the simulator or virtual hardware environment, which can simulate different hardware input signals and transmission data to identify potential hardware problems~\cite{ISERMANN19981}. This method is particularly effective in chip internal logic verification, bus communication testing, and data processing.

\section{Conclusion}
In this work, we present an in-depth review and summary of fuzzing test technology based on LLMs, and cover a wide range of applications in AI and non-AI software fields. We summarize the frameworks and principles of different LLM-based fuzzers and discuss how these technologies introduce LLMs to enhance traditional fuzzing test technologies. Compared with traditional fuzzers, the LLM-based fuzzers provide superior API and code coverage to find more complex bugs and improve the automation of fuzzing tests. In this way, LLM-based fuzzing technology demonstrates great potential in advancing the field of software testing. The insights and results of this work are expected to be a valuable resource for researchers and practitioners in the field, guiding the development of more efficient, reliable, and automated fuzzing test methods by using the LLMs.

\balance
\bibliographystyle{IEEEtran}
\bibliography{aaai25}

\end{document}